\documentclass{article}
\usepackage{spconf, amsmath, graphicx}
\usepackage{indentfirst}
\usepackage{float}
\usepackage{graphicx}
\usepackage{caption}
\usepackage{subcaption}
\usepackage{adjustbox}
\usepackage{siunitx}
\usepackage{color}
\usepackage{amsfonts}
\usepackage{booktabs}
\def\ps@IEEEtitlepagestyle{%
	\def\@oddfoot{}%
	\def\@evenfoot{\mycopyrightnotice}%
}
\def\mycopyrightnotice{%
	{\footnotesize This paper is accepted to be published in: 2018 IEEE International Conference on Image Processing, Oct 7-10, 2018, Athens, Greece. \par \footnotesize IEEE Copyright Notice: \copyright IEEE 2018 Personal use of this material is permitted. Permission from IEEE must be obtained for all other uses, in any current or future media, including reprinting/republishing this material for advertising or promotional purposes, creating new collective works, for resale or redistribution to servers or lists, or reuse of any copyrighted component of this work in other works.
\hfill}
	\gdef\mycopyrightnotice{}
}
\newcommand\blfootnote[1]{%
	\begingroup
	\renewcommand\thefootnote{}\footnote{#1}%
	\addtocounter{footnote}{-1}%
	\endgroup
}

\title{Nonlinear shape regression for filtering segmentation results from calcium imaging
}
\name{Jie Wang$^{\dagger}$, Zhongxiao Fu$^{\ddagger}$, Nasrin Sadeghzadehyazdi$^{\dagger}$, Jonathan Kipnis$^{\ddagger}$and Scott T. Acton$^{\dagger}$}
\address{$^\dagger$Department of Electrical \& Computer Engineering and $^\ddagger$Department of Neuroscience,\\University of Virginia, Charlottesville, VA 22904, USA}

\begin{document}

\maketitle

\begin{abstract}
A shape filter is presented to repair segmentation results obtained in calcium imaging of neurons \textit{in vivo}. This post-segmentation algorithm can automatically smooth the shapes obtained from a preliminary segmentation, while precluding the cases where two neurons are counted as one combined component. The shape filter is realized using a square-root velocity to project the shapes on a shape manifold in which distances between shapes are based on elastic changes. Two data-driven weighting methods are proposed to achieve a trade-off between shape smoothness and consistency with the data. Intuitive comparisons of proposed methods via projection onto Cartesian maps demonstrate the smoothing ability of the shape filter. Quantitative measures also prove the superiority of our methods over models that do not employ any weighting criterion.
\end{abstract}
\vspace{-0.2cm}
\blfootnote{\mycopyrightnotice}
\section{Introduction}
\vspace{-0.3cm}
\label{sec:intro}
The mammalian brain is a complex system, mainly due to the large number of neurons that form complicated interconnected networks. Human brains contain almost 100 billion neurons, and mouse brains have about 75 million neurons \cite{azevedo2009equal}. Each of the neurons can receive thousands of inputs and can innervate thousands of downstream neurons. Since brain functions require coordinated activation of large groups of neurons across different brain regions, recording and understanding neural activity at circuit level are fundamental for understanding brain cognitive functions. 

In the 1950s, Hubel and Wiesel discovered links between visual stimuli and neurons in the visual cortex using implanted electrodes that monitor single cells. The current revolution in calcium imaging is accommodating such observation of neural function for a multitude of neurons without invasive electrodes. Calcium imaging with fluorescent proteins provides a convenient optical way to record large neuron populations with precision down to the single action potential level \cite{chen2013ultrasensitive}. With progress from two parallel areas, calcium sensitive fluorescent protein engineering \cite{chen2013ultrasensitive,dana2016sensitive} and calcium imaging techniques \cite{ghosh2011miniaturized,ahrens2013whole,zong2017fast}, \textit{in vivo} calcium imaging has become a standard method to investigate neural circuit mechanisms underlying cognitive behaviors. However, new calcium imaging methods, in contrast to previous sparse neuron and short-term imaging recordings, generate a huge volume of data with hundreds and thousands of neurons resulting from several hours of recording. To analyze these large datasets, high throughput automated image analysis methods, such as advanced image segmentation techniques, are required to not only be able to identify single neurons, but also to detect single calcium events.

Over the years, researchers have explored numerous image segmentation techniques for the analysis of biomedical images. Existing segmentation techniques \cite{acton2009biomedical} such as thresholding, edge detection, active contours \cite{mansouri2004constraining,Mukherjee2015} and morphological methods \cite{acton2000area,bosworth2003morphological} are still limited to single-neuron analysis.  Recently, efforts in the image analysis research community have attempted to avoid the identification of multiple components as a single component by using graph cut method  \cite{browet2016cell} and an iterative local level set evolution \cite{Wang2017Bact3D}. Both approaches allow identification of individual cells; however, the seeded segmentation strategy in both methods is challenging in our datasets given noisy signals from non-activated neurons. Also, the \textit{in vivo} videos increase the difficulty of locating seeds in individual cells. There are also some post-segmentation algorithms that attempt to isolate connected components, including the watershed transform \cite{yang2014automatic} and concavity identification \cite{kong2011partitioning}. The former algorithm fails in the inhomogeneity present in our datasets, while the latter method suffers from the rough boundaries extracted from preliminary segmentation.

\begin{figure*}[t]
	\centering
	\includegraphics[width=0.86\textwidth]{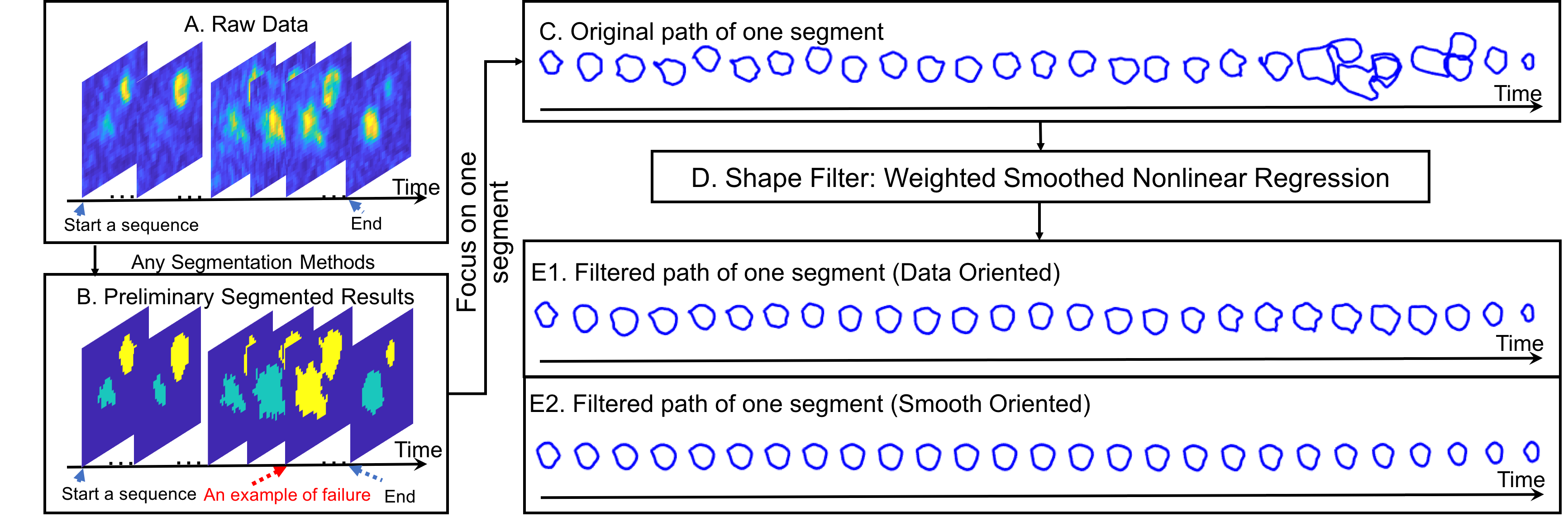}
	\vspace{-0.3cm}
	\caption{Flow chart of the shape filtering approach.}
	\label{fig:flowchart}
	\vspace{-0.6cm}
\end{figure*}

In this paper, we explore a novel perspective, a post-segmentation shape filter, to transform the preliminary segmentation produced on a temporal sequence of shapes. It provides a smooth shape evolution path targeting on single neuron activity along time by filtering out the unnecessary and incomplete segments and smoothing the boundaries of preliminary results. To realize this method, a modified smooth nonlinear regression (spline) of shapes on a Riemannian manifold is proposed based on the shape space introduced in \cite{Anuj2011SRVD}. Although several works \cite{baust2015total,su2012fitting, hinkle2014intrinsic,kuhnel2017stochastic} proposed smooth splines for regression on a manifold by evaluating polynomial energy-minimization functions, they are not able to automatically detect and exclude the non-smooth shapes while preserving the structure of the target object. \cite{Liu2017face,Cao2012ShapeRegression} use deep learning on shape regression for face alignment. These methods are not able to split connected components and are not suitable for a limited dataset.
\vspace{-0.3cm}

\section{shape filter}
\label{sec:shaperep}
\vspace{-0.3cm}
Due to the limitations of the segmentation methods, some of the results show separate cells as one single connected component. In this paper, we propose a novel method to automatically filter out such outliers and then provide a smooth segmentation of the cell of interest from the rest of the connected component. We design a shape filter that generates a smooth path on a manifold to represent shape evolution through time. Fig. \ref{fig:flowchart} shows the flow chart of the shape filter framework. In the following sections, we present a detailed description of a locally weighted shape smooth regression method that provides a smooth segmentation based on the shape evolution path fitted to the data. 
\vspace{-0.6cm}
\subsection{Data preparation and shape representation}
\vspace{-0.2cm}
As shown in Fig. \ref{fig:flowchart}, the input of the shape filter is the preliminary result after any segmentation. Each time, only one neuron is chosen to be analyzed. Then, the shape information is extracted and represented by equidistant points located on the boundary of the segmented body in the Cartesian coordinates (see Fig. \ref{fig:flowchart}C). We use the square-root velocity (SRV) representation of the shapes \cite{Anuj2007SRV,Anuj2011SRVD} in order to compare the shapes extracted from the time-indexed segmented data. Under the SRV representation, each shape is defined on a Riemannian manifold $\mathcal{M}$  with locally assembled SRV Euclidean coordinates $q$, defined as follows:
\vspace{-0.3cm}
\begin{equation}
 \vspace{-0.3cm}
q(s) = \frac{\dot{\beta(s)}}{\sqrt{||\dot{\beta(s)}||}}
\end{equation}
where $\beta(s)$ represents the shape, which is a closed curve parameterized by equal distant arclength $s\in [0,2\pi]$. $\dot{\beta(s)}$ shows the gradient of this curve and $||\cdot||$ denotes the Euclidean norm. Suppose $\alpha: \mathbb{R} \rightarrow \mathcal{M}$ is the geodesic path between two arbitrary shapes on the defined manifold. Then, the difference of two shapes can be evaluated using the geodesic distance between the two SRV-transformed shapes:
\vspace{-0.2cm}
\begin{equation}
 \vspace{-0.3cm}
d_{g}  = \int_{0}^{1}\sqrt{\langle\dot{\alpha(t)},\dot{\alpha(t)}\rangle}dt
\end{equation}
where $\alpha(0)$ and $\alpha(1)$ represent the initial and final positions of the path respectively. $\dot{\alpha(t)}$ shows the gradient of this geodesic path. When $dt$ is infinitesimally small, the geodesic distance is approximately the integral length of the gradient magnitudes.
\vspace{-0.3cm}
\subsection{Weighted smoothed nonlinear regression}
\vspace{-0.2cm}
After transforming the shape data into the SRV representation, an optimization problem is solved to filter out the extra components (\textit{outliers}) in the preliminary result by interpolating the likely smooth shapes along the time-indexed calcium firing path. At the same time, the filtered path can also smooth the boundary of segmentation results by imposing several constraints. 

We use $\phi: \mathbb{R} \rightarrow \mathcal{M}$ to denote the original evolution path and $\gamma: \mathbb{R} \rightarrow \mathcal{M}$ to represent the filtered new path on the manifold. Both of these two paths are approximately differentiable and have the geodesic distance defined as aforementioned. Then, $\gamma$ is estimated by optimizing the following regression problem, which is modified from De Boor's approach \cite{Deboor2001Spline} by proposing automatic outlier detectors ($w$) on the space of a shape manifold:
 \vspace{-0.2cm}
\begin{equation}
 \vspace{-0.2cm}
\label{eq:main}
\small{\min_{\gamma} \rho\underbrace{\sum_{t=start}^{end}w(t)|\phi(t)-\gamma(t)|^{2}}_{data\ term}+(1-\rho)\underbrace{\int|\mathit{D}^{2}\gamma(t)|}_{smoothness\ term}}
\end{equation}

This modified shape regression model aims to find a desired minimizer $\gamma$ that balances the trade-off between approaching the original data and smoothing the filtered path (using MATLAB function csaps). In equation (\ref{eq:main}), $t$ is time index for each shape, which is also used in the following sections in this paper. $\mathit{D}^{2}\gamma(t)$ is the second derivative of path $\gamma$, which characterizes the changes of shapes along the fitted path. $\rho\in[0,1]$ is the smoothing parameter that reflects the emphasis on data or smoothness. Note that, when $\rho$ is close to 1, $\gamma$ becomes the spline of the input data $\phi$; and when $\rho$ becomes small, $\gamma$ will be smoother with fewer shape changes along time in terms of SRV transformed shape representation. $w(t)$ is the set of local weights used to estimate the outliers automatically, which will be explained in details in section \ref{ssec:w}.
\vspace{-0.3cm}
\subsection{Local weight selections}
\vspace{-0.2cm}
\label{ssec:w}
To detect clutter found in the segmentation of the calcium images, we propose a locally weighted shape regression strategy. Four weight selection methods are discussed in this section, where the third and the fourth are the proposed local weights for the shape filter. 

(1) \textit{Unity weighting}: $\mathbf{w_{1}} = \mathbf{1}$ is a column vector with all the values equal to 1, which tends to no preference in the weighting of segments.

\begin{figure}[h]

	\centering
	\setlength{\tabcolsep}{0.03cm}
	\begin{tabular}{ccc}
		\includegraphics[width = 0.33\linewidth,height=0.27\linewidth]{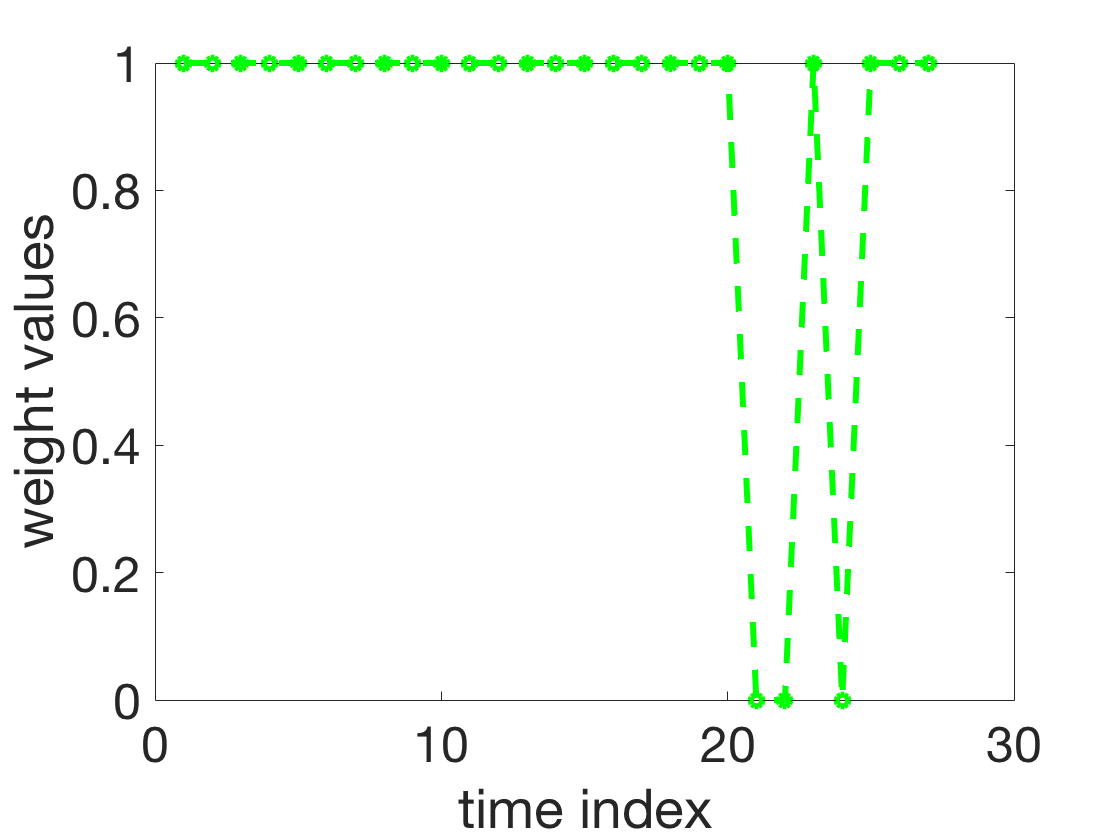} &
		\includegraphics[width = 0.33\linewidth,height=0.27\linewidth]{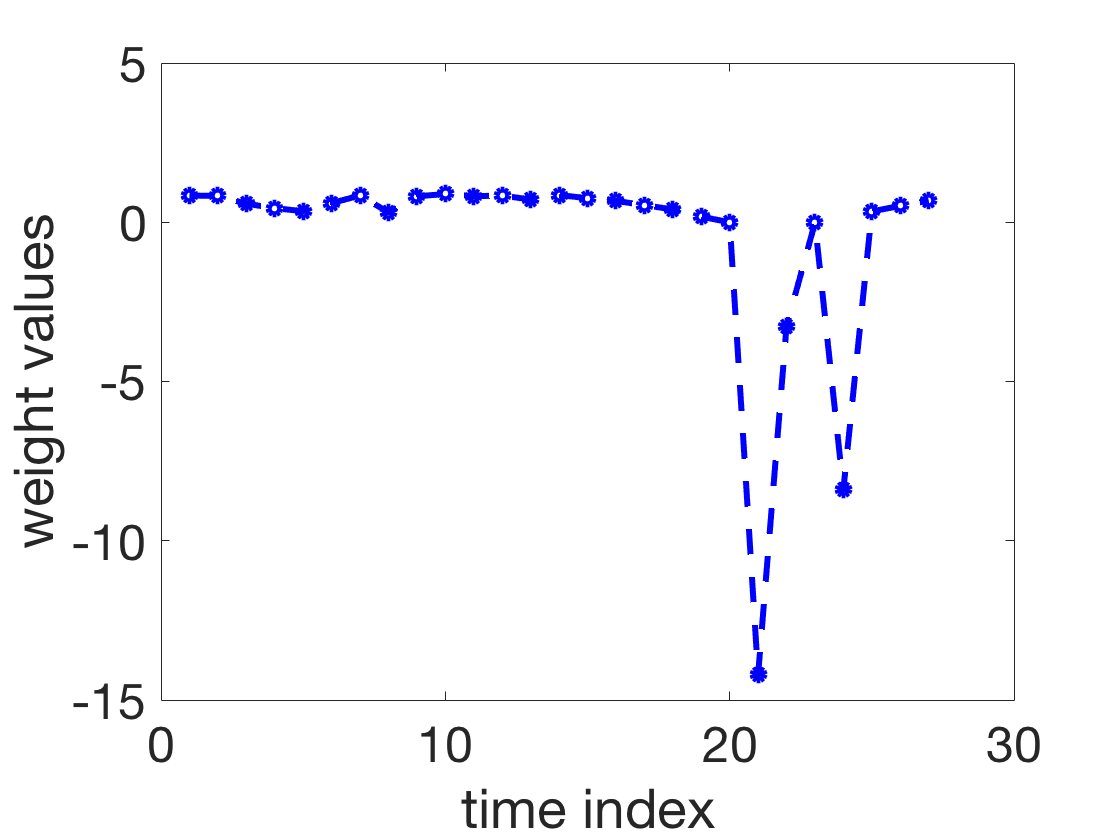} &
		\includegraphics[width = 0.33\linewidth,height=0.27\linewidth]{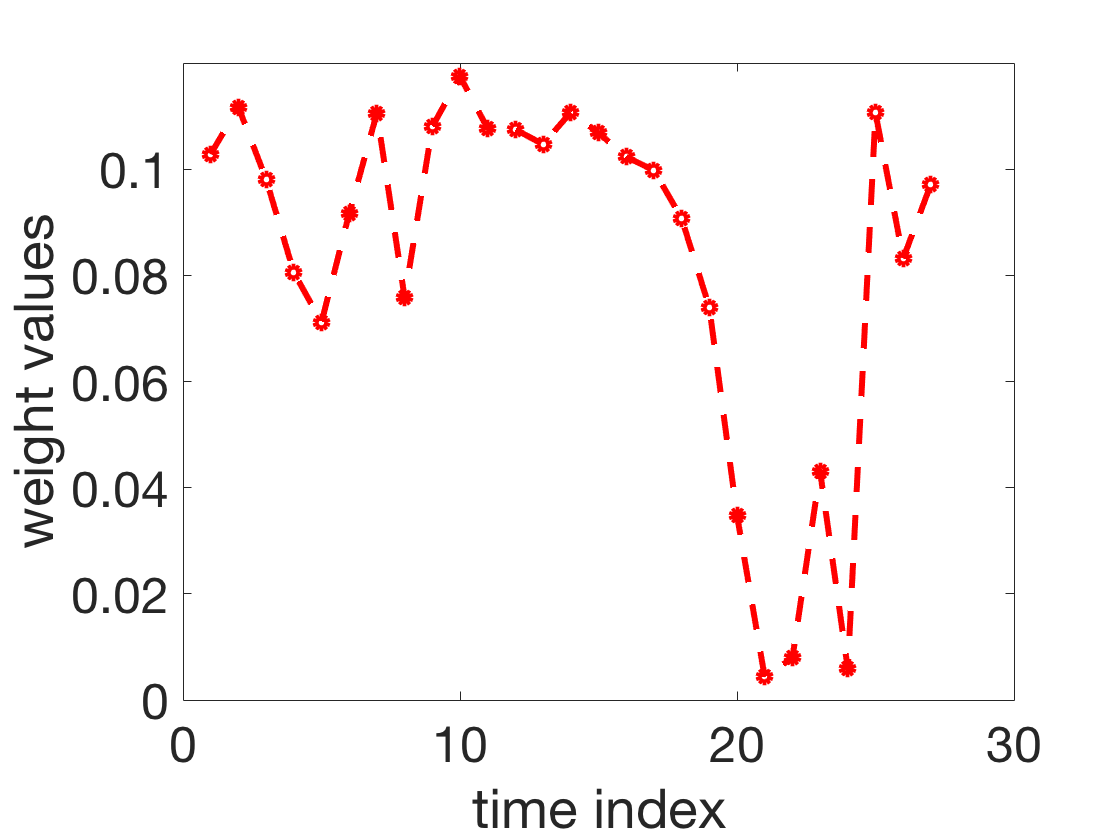} \\
		\small{(i) Piecewise Constant} & \small{(ii) Bi3 local} & \small{(iii) sGaussian}
	\end{tabular}
	\vspace{-0.3cm}
	\caption{Three weighting responses of a sample input. All of them can modify weights based on the local shape information. However, (i) needs additional analysis to detect the \textit{outliers}; (ii) and (iii) are instantaneous and automatic data-driven weightings. }
	\vspace{-0.3cm}
	\label{fig:w}
\end{figure}

(2) \textit{Piecewise constant weighting}: The second weighting scheme is given by $w_{2}(t)$, which is a step function (Fig. \ref{fig:w}(i)). 
\vspace{-0.2cm}
\begin{equation}
\vspace{-0.2cm}
w_{2}(t) = \left\{
\begin{aligned}
C, &\text{  if not outliers}\\
0,  &\text{  otherwise}
\end{aligned}
\right.
\end{equation}
Any constant $C$ can be chosen based on the desire. For those outliers, which need additional steps (out of the main regression problem) to identify, original data will be ignored with data term equals to zero.

(3) \textit{Bi3 local shape weighting}: Inspired by tricube kernel model and robust local regression model in \cite{altman1992tricube}, 
Bi3 local shape weight is defined as: 
 \vspace{-0.3cm}
\begin{equation}
\label{eq:bi3}
w_{3}(t) = A*(1-(\frac{r(t)}{\mathrm{median}(r(t))+\tau})^{3})^{3}
 \vspace{-0.3cm}
\end{equation}
where
\begin{equation}
r(t) = d_{g}(\delta(t), \phi(t))
 \vspace{-0.2cm}
\end{equation}
\begin{equation}
 \vspace{-0.2cm}
\sigma_r = \frac{1}{N}\sum_{t=1}^{N}(|r(t)-median(r(t))|)
\end{equation}
\begin{equation}
 \vspace{-0.2cm}
\tau = \sigma_r +( \sigma_r -min(r(t)))
\end{equation}
where $r(t)$ is the residual geodesic distance from the true data to a fitted smooth spline $\delta: \mathbb{R} \rightarrow \mathcal{M}$ with a small weight for the data term in equation (\ref{eq:main}). $\sigma_r$ specifies the mean deviation of residuals from $\delta$. $median(r(t))$ is treated as mean shape in the sequence instead of the traditional mean because the outliers will affect this value. $\tau$ is the tolerance for residual deviations. $\tau$ will automatically assign a negative feedback to data term when the data is far larger than tolerance from the smoothed spline $\delta$. The formula to calculate this value is inspired by the skew measure in statistical analysis. $A$ is a constant that can amplify the proportion of the data term. The cubic weight is capable of yielding wider range for data when the base of inner cube is closer to zero \cite{altman1992tricube}, which increases the acceptance of shapes that is similar to the mean shape. 


(4) \textit{Modified shape Gaussian (sGaussian) weighting}: The Gaussian model is also a popular weighting selection in local regression problems, such as in \cite{loader2006Gaussian}. To adapt the framework of Gaussian weighting to shapes, the weighting may be computed using:
\vspace{-0.3cm}
\begin{equation}
\vspace{-0.3cm}
w_{4}(t)= \frac{1}{\sqrt{2\pi\sigma^{2}}}\exp(-\frac{d_{g}^{2}(\phi(t),q_{median})}{2\sigma^{2}})
\end{equation}
where 
\begin{equation}
 \vspace{-0.2cm}
\sigma^{2}= \frac{1}{N}\sum_{t=1}^{N}(d_{g}(\phi(t),q_{median}))^{2}
\end{equation}
Here, $q_{median}$ represents the Euclidean median shape along the input path $\alpha$ with $N$ shapes in the SRV representation. Moreover, Euclidean distance in the normal Gaussian cases is replaced by the geodesic distance. The weighting response according to a sample input datum is shown in Fig. \ref{fig:w}(iii).
 \vspace{-0.3cm}
\vspace{-0.4cm}
\section{Experimental Results}
\vspace{-0.2cm}
The analysis and comparison of the methods examined here are performed on 10 videos with $0.065$ s time sampling. Each frame in the videos has $512\times512$ resolution with $\SI{0.9}{\micro\metre}\times \SI{0.9}{\micro\metre}$ pixels. Activated neurons observed by calcium imaging have diameters that vary from $\SI{10}{\micro\metre}$ to $\SI{20}{\micro\metre}$. For experimental data, we focused on 20 regions of interest with 10 to 40 frames per sample, where neighboring neurons are activated at the same time. In these cases, separating different neurons to individual components is a challenging task. We used basic thresholding and morphological methods to produce the preliminary segmentation results. Then, we compared the experimental results on these datasets using \textit{unity weighting}, \textit{Bi3 local shape weighting} and \textit{modified shape Gaussian weighting} quantitatively and qualitatively. 

\begin{table}[b]
	\vspace{-0.5cm}
	\centering
	\caption{Comparison of different weights.}
	\vspace{-0.3cm}
	\label{tbl:results}
	\begin{tabular}{@{}cccc@{}}
		\cmidrule(l){2-4}
		& Bi3 local      & sGaussian      & Unity          \\ \midrule
		Dice & \textbf{0.918} & 0.915          & 0.802          \\
		MSE  & 0.016          & \textbf{0.010} & 0.226          \\ \bottomrule\end{tabular}
	\vspace{-0.3cm}
\end{table}
\begin{figure*}[t]
	\centering
	\begin{tabular}{cc}
		\includegraphics[width = 0.49\textwidth,height = 0.46\textwidth]{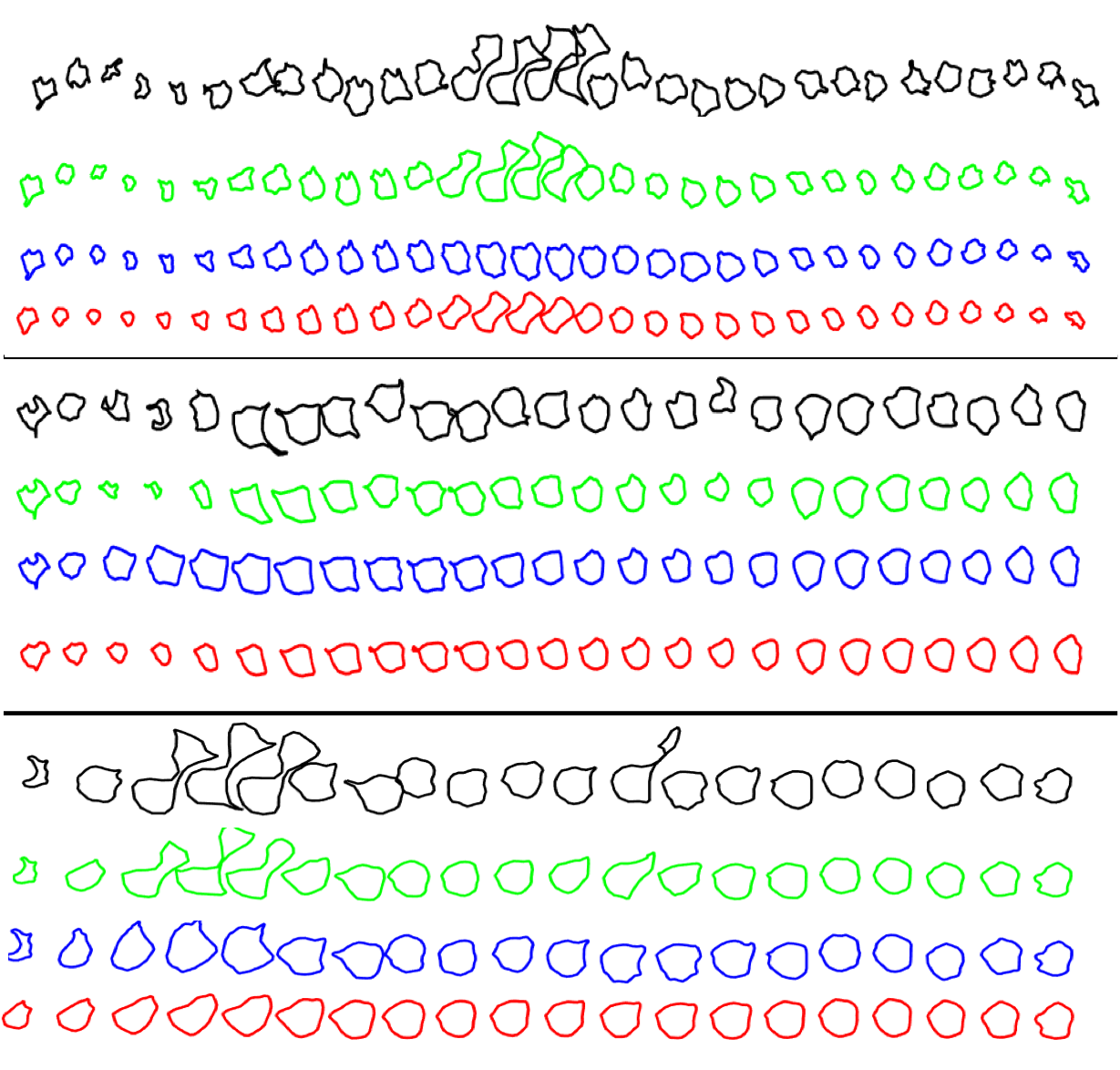} &
		\includegraphics[width = 0.49\textwidth,height = 0.46\textwidth]{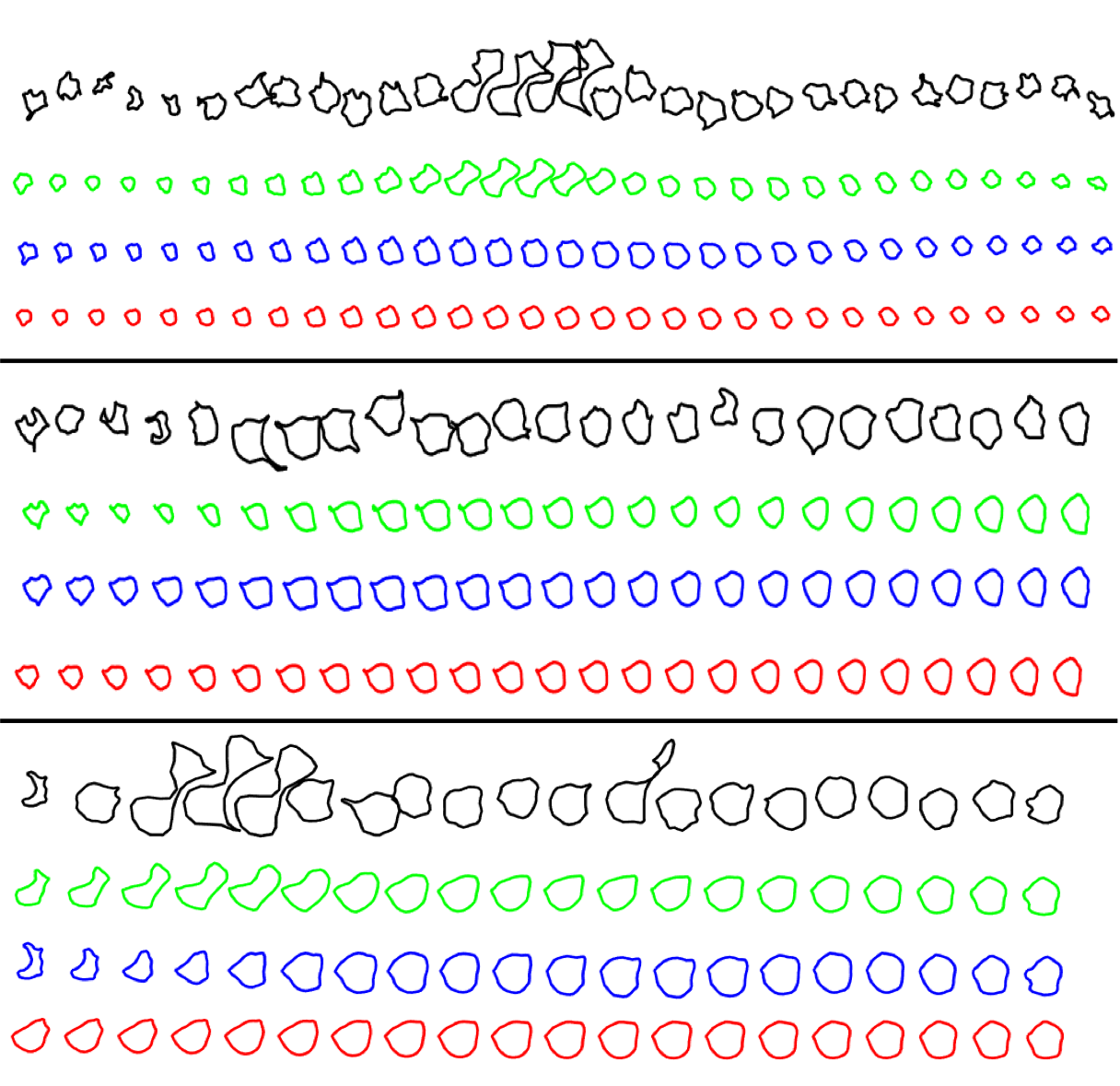}\\
		\small{Data oriented} & \small{Smoothness oriented}
	\end{tabular}
	\vspace{-0.3cm}
	\caption{Comparison of data and smoothness oriented results. In each section: 1st row:  input path;  2nd row: filtered path using unity weight; 3rd row: filtered path using Bi3 local; 4th row: filtered path using sGaussian. The first column shows data oriented results with larger $\rho$ values and the second column shows smoothness oriented results with smaller $\rho$.}
	\label{fig:r}
	\vspace{-0.2cm}
	\vspace*{\floatsep}
	\centering
	\begin{tabular}{cccccc}
		\includegraphics[width = 0.147\textwidth]{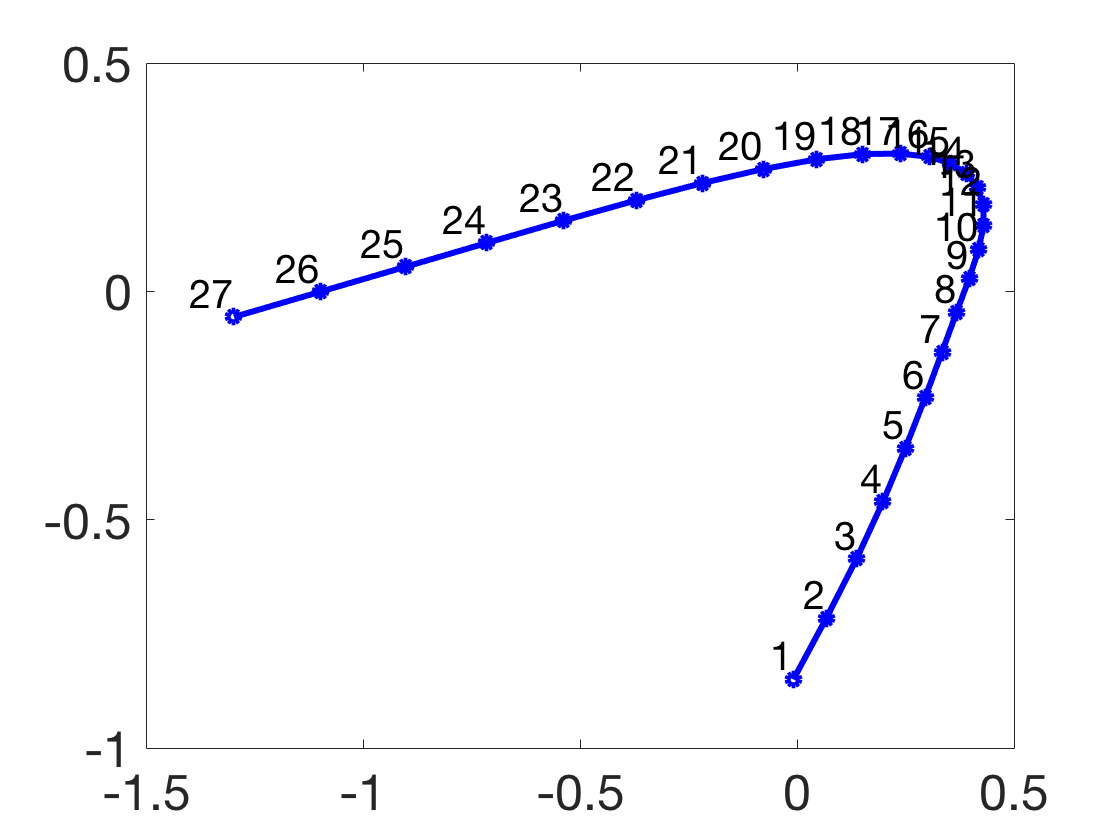} &
		\includegraphics[width = 0.147\textwidth]{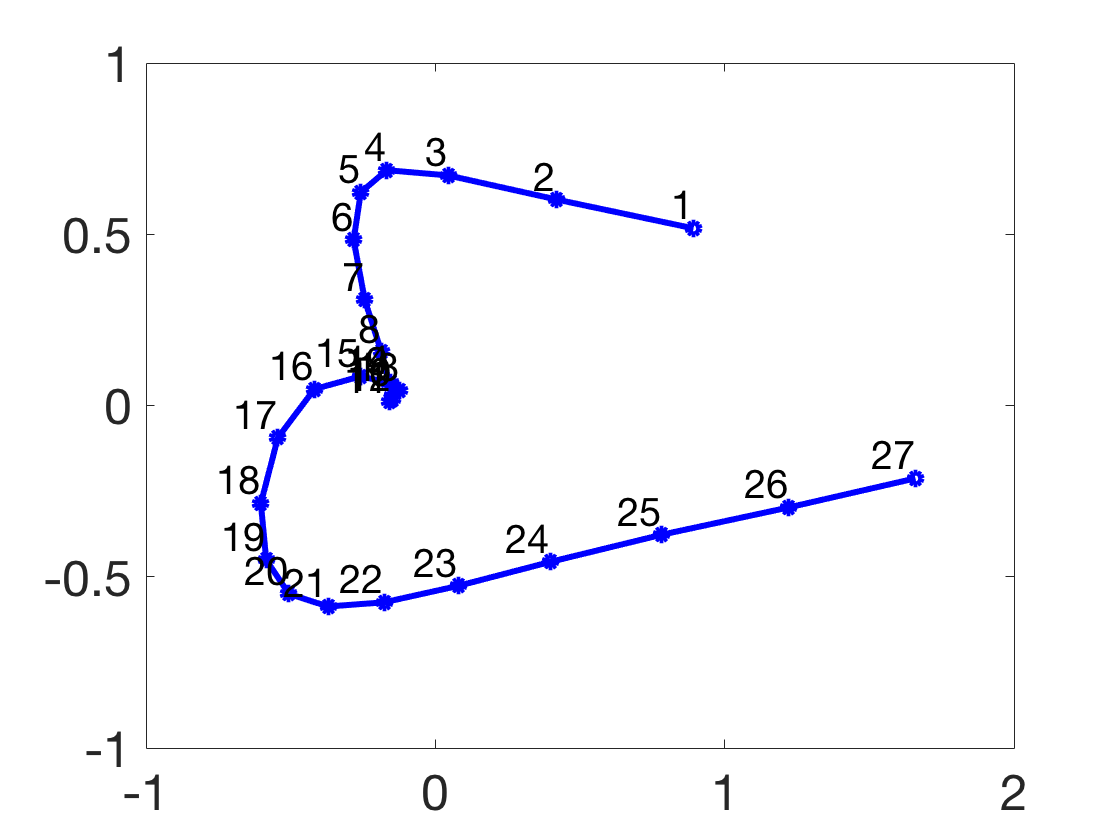} &
		\includegraphics[width = 0.147\textwidth]{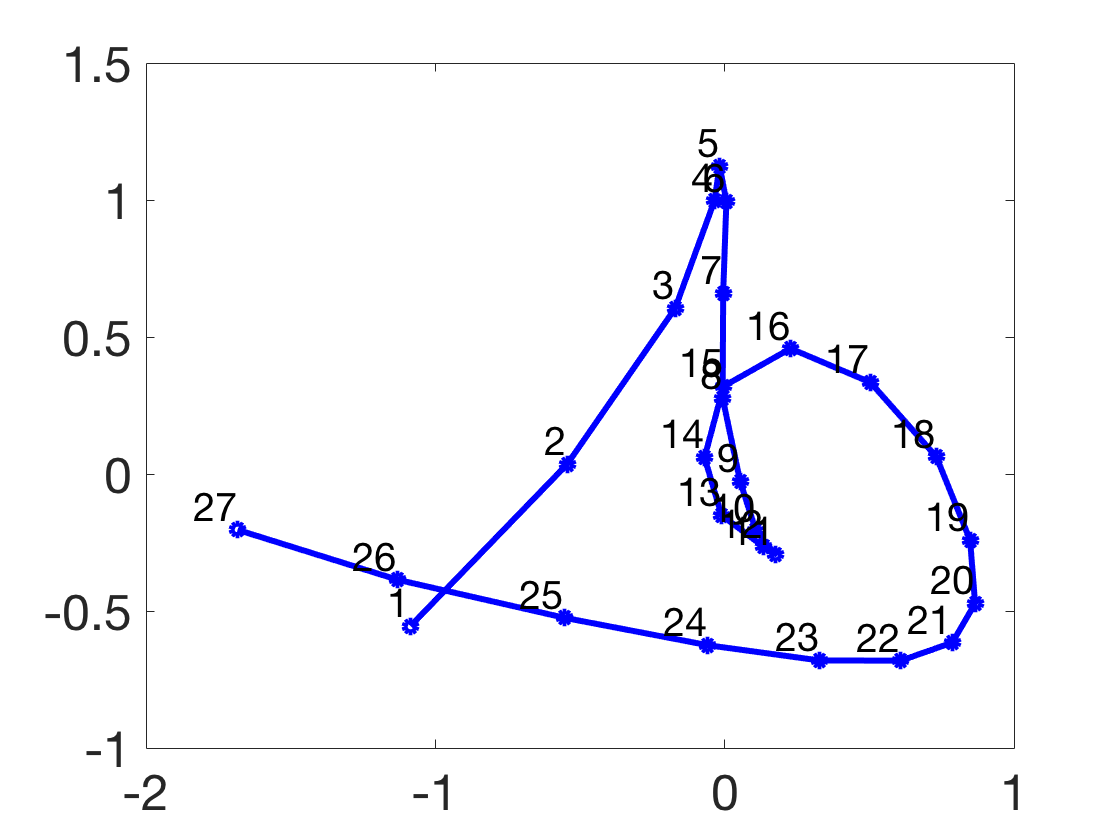} &
		\includegraphics[width = 0.147\textwidth]{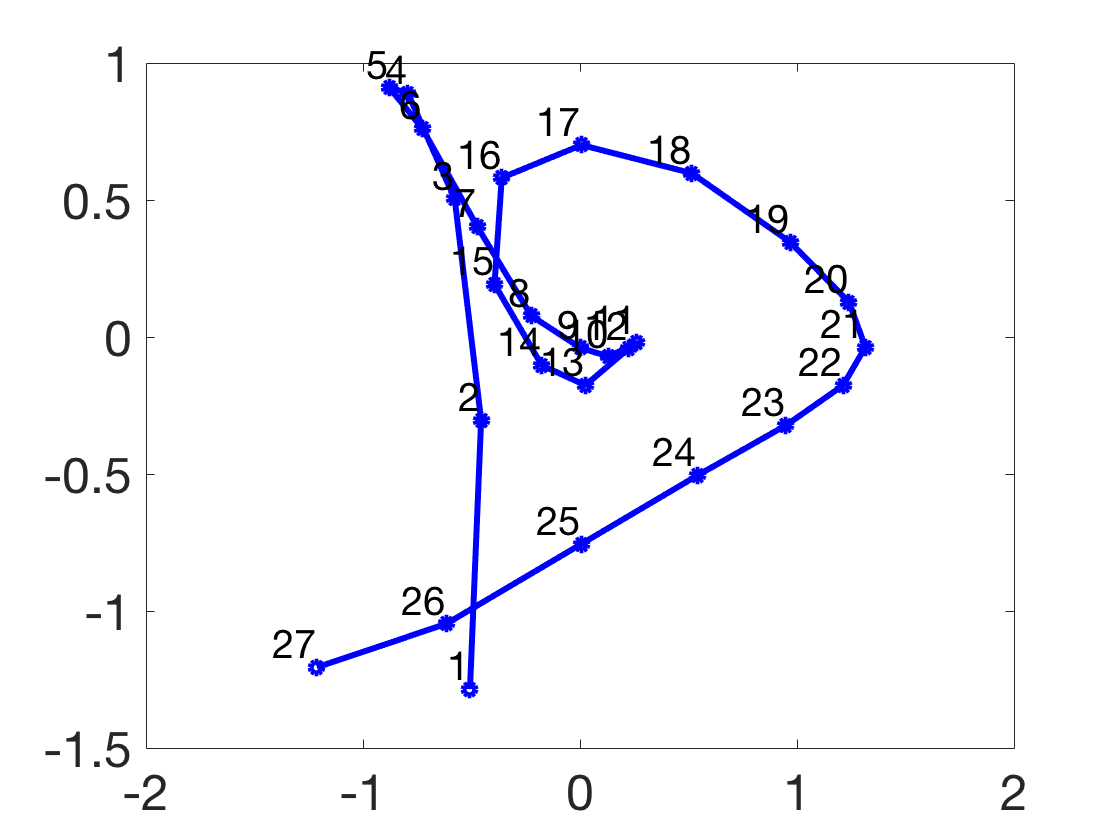} &
		\includegraphics[width = 0.147\textwidth]{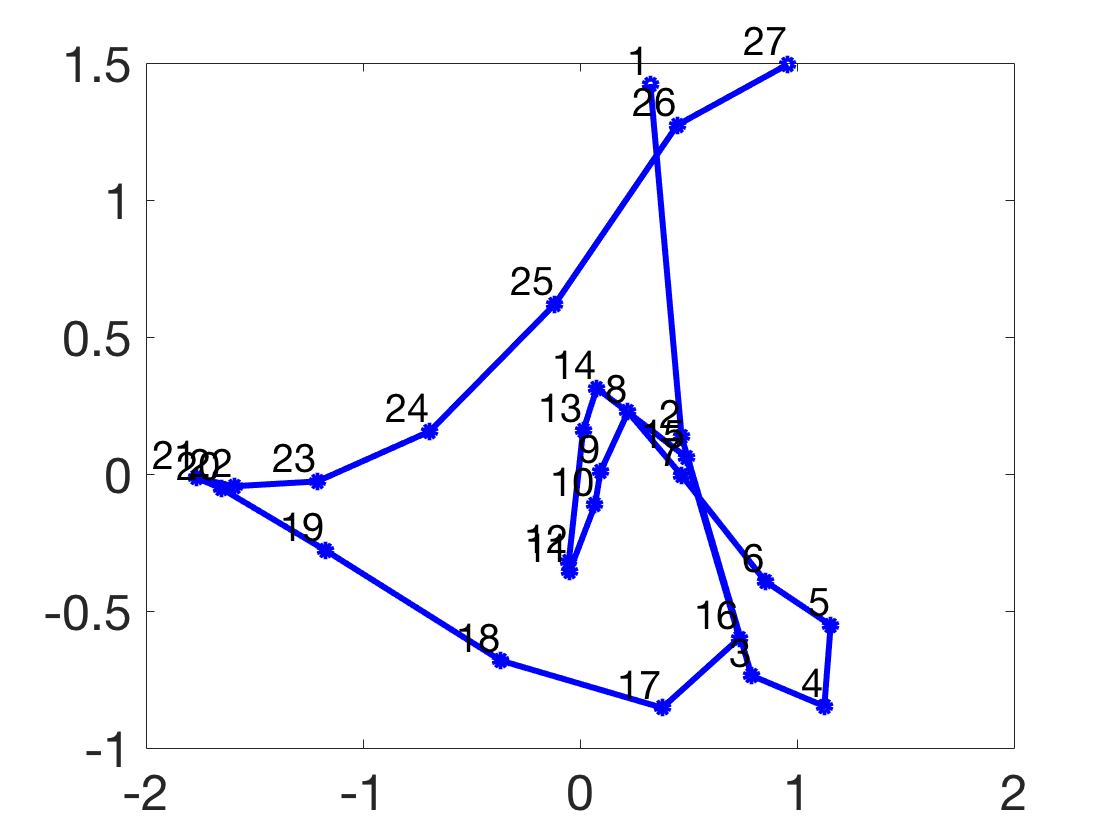} &
		\includegraphics[width = 0.147\textwidth]{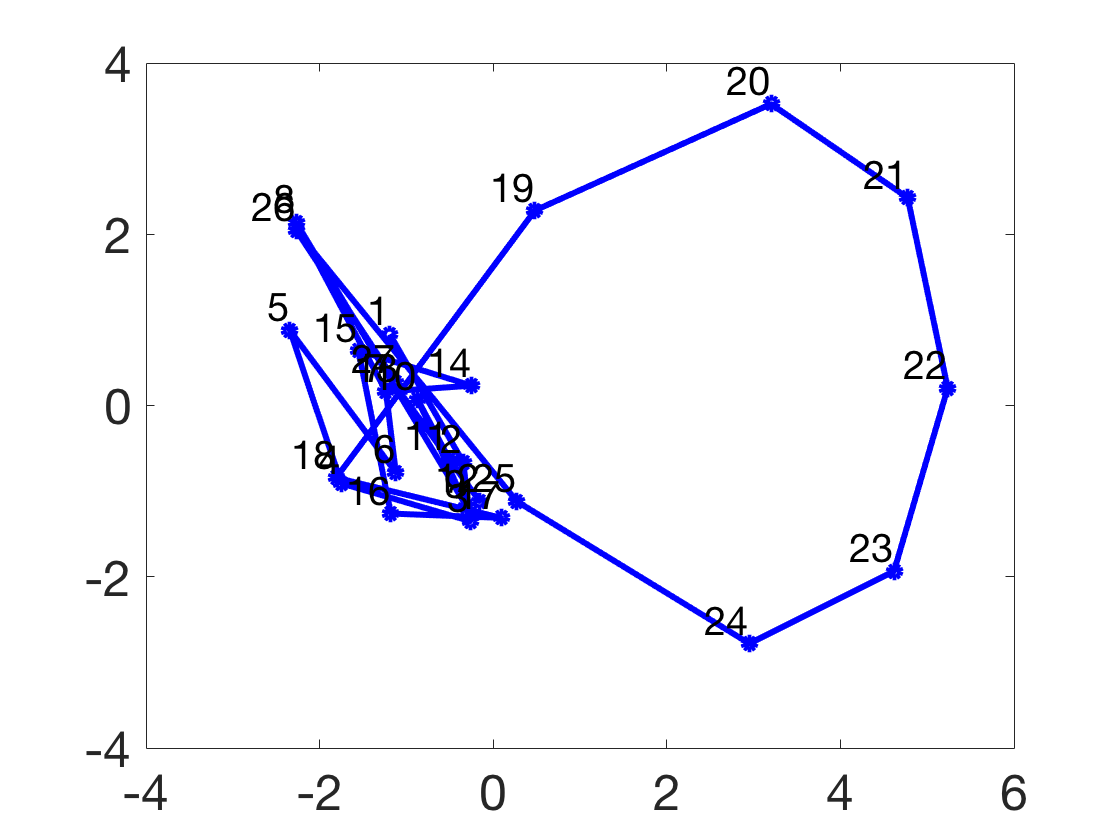} \\
		\includegraphics[width = 0.147\textwidth]{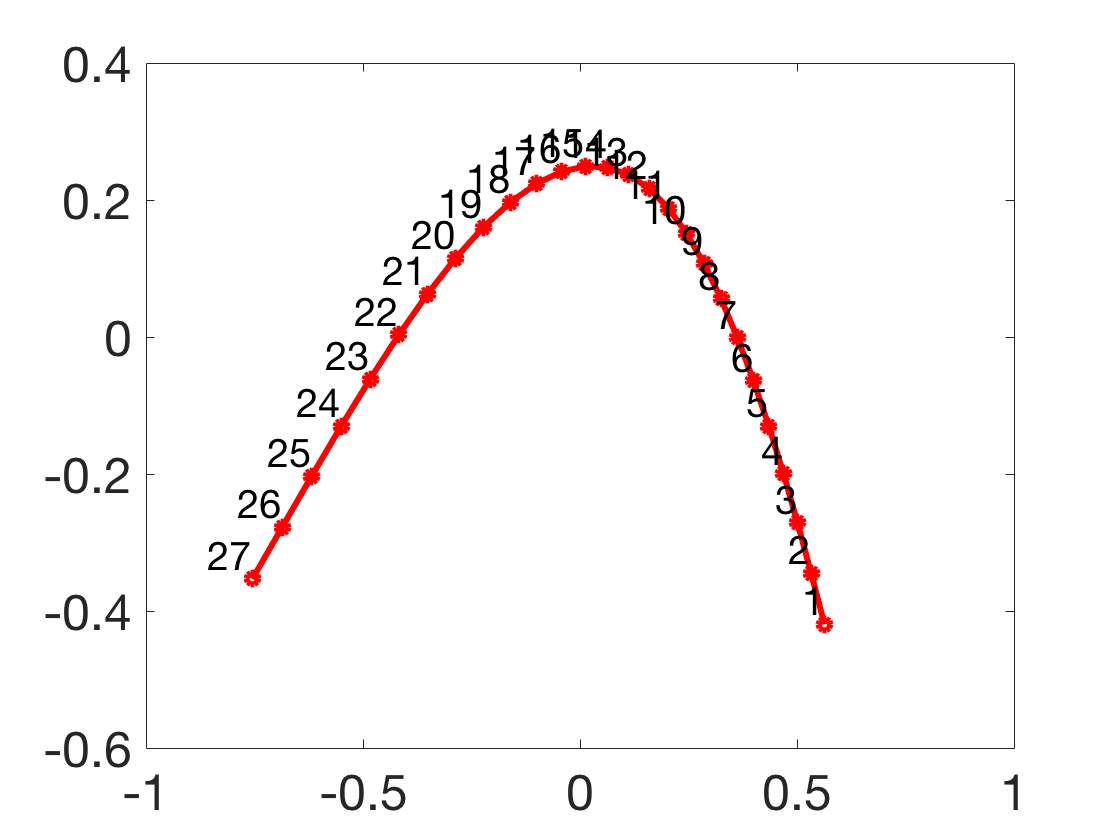} &
		\includegraphics[width = 0.147\textwidth]{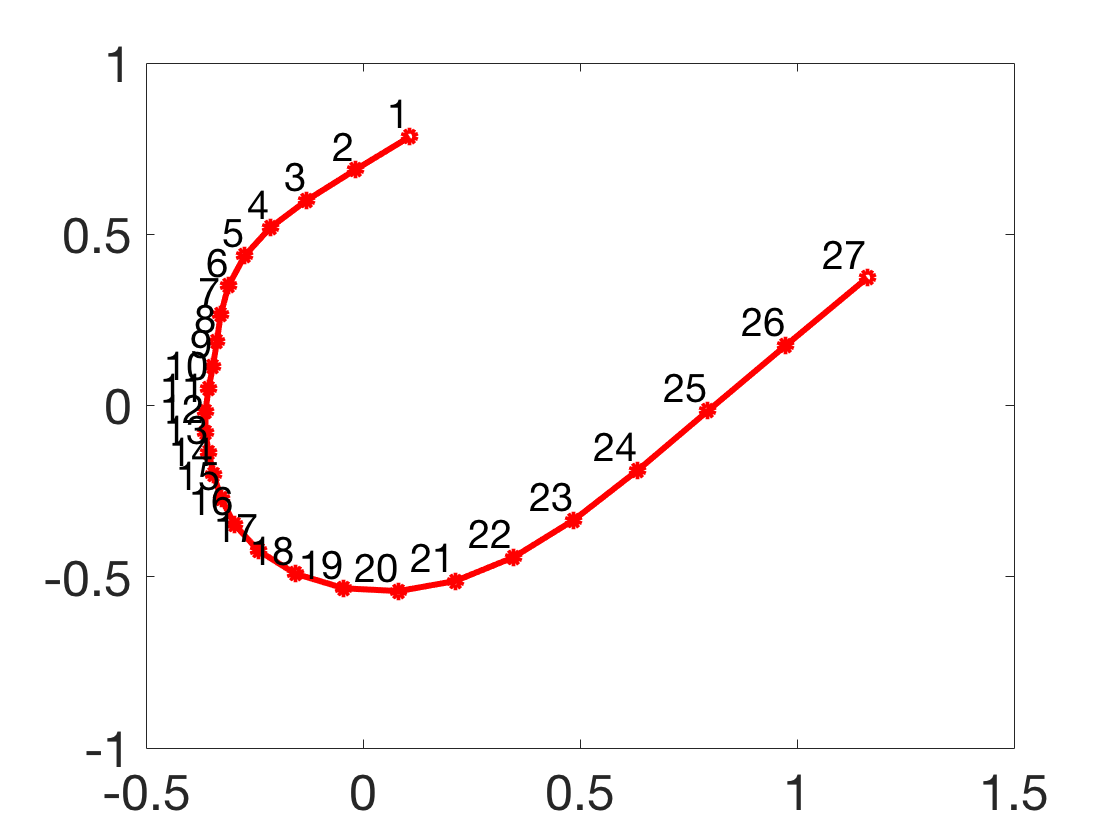} &
		\includegraphics[width = 0.147\textwidth]{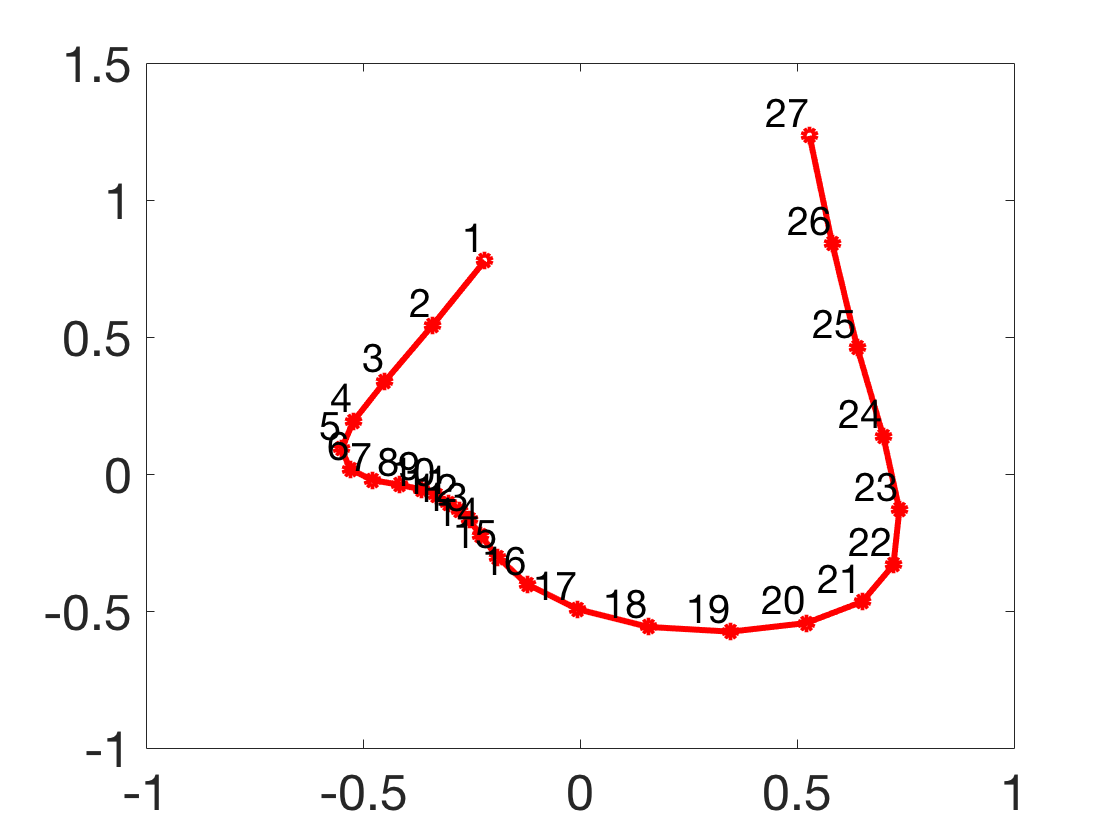} &
		\includegraphics[width = 0.147\textwidth]{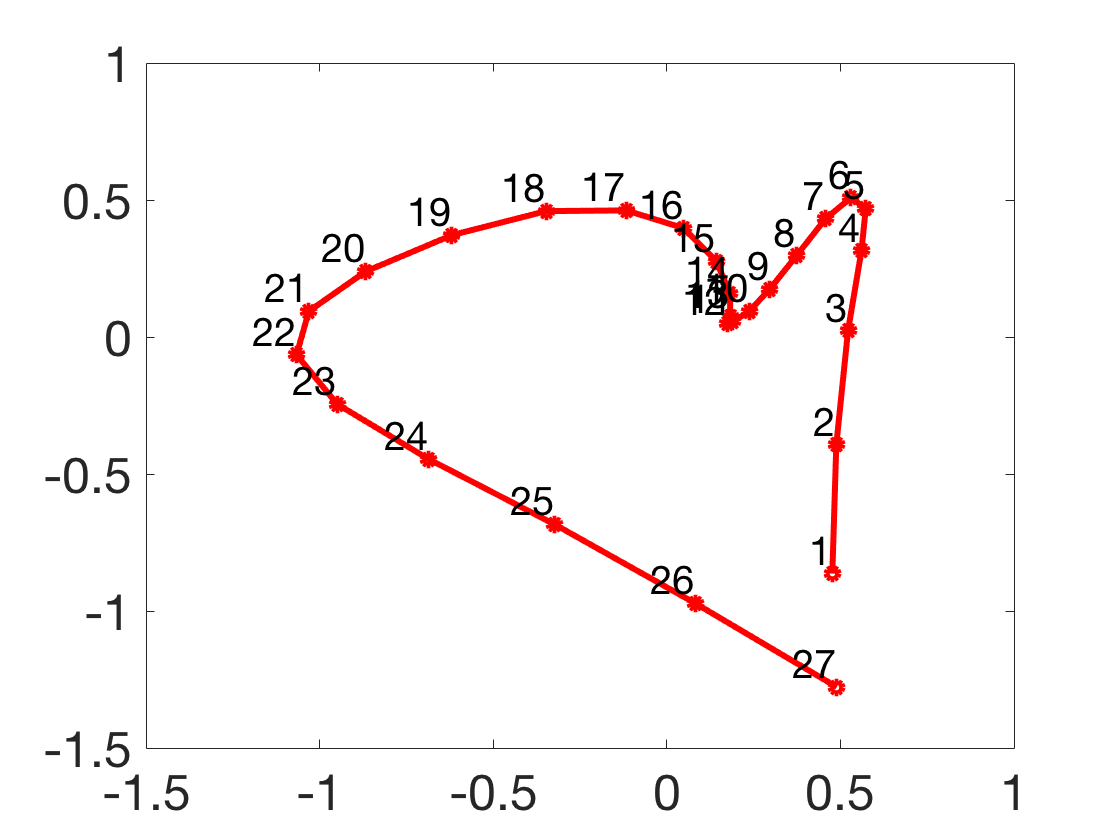} &
		\includegraphics[width = 0.147\textwidth]{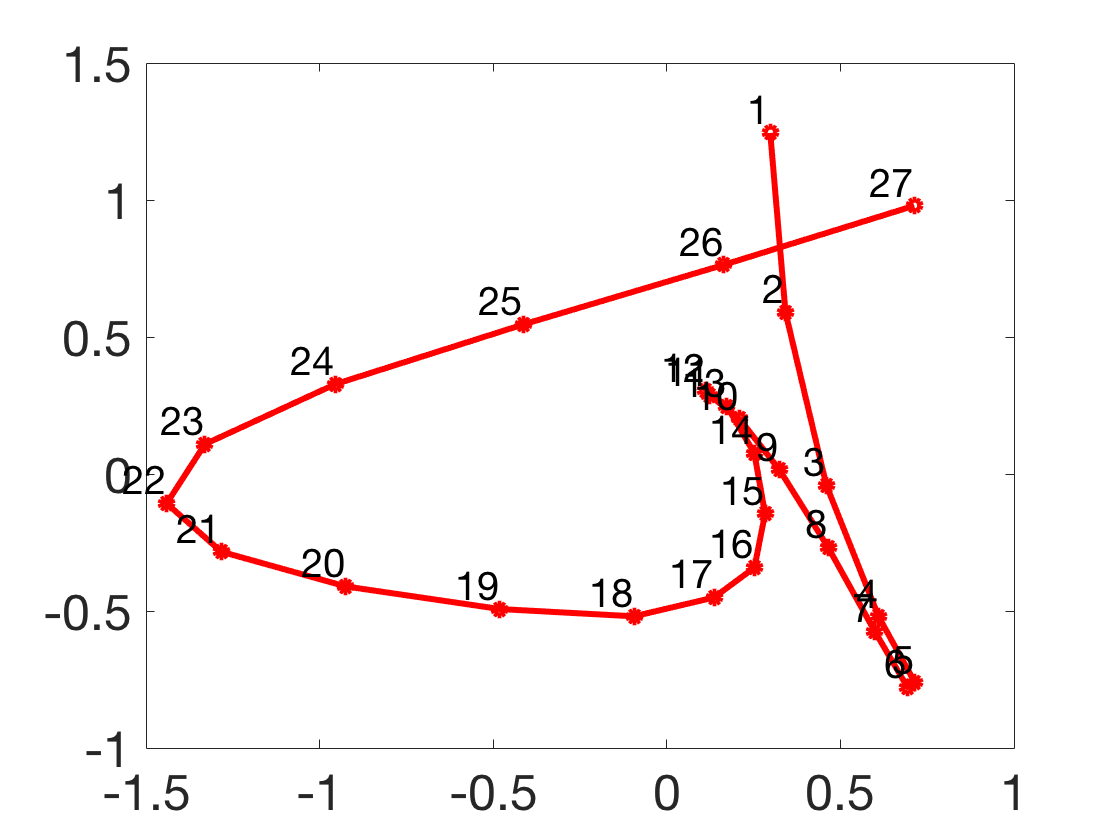} &
		\includegraphics[width = 0.147\textwidth]{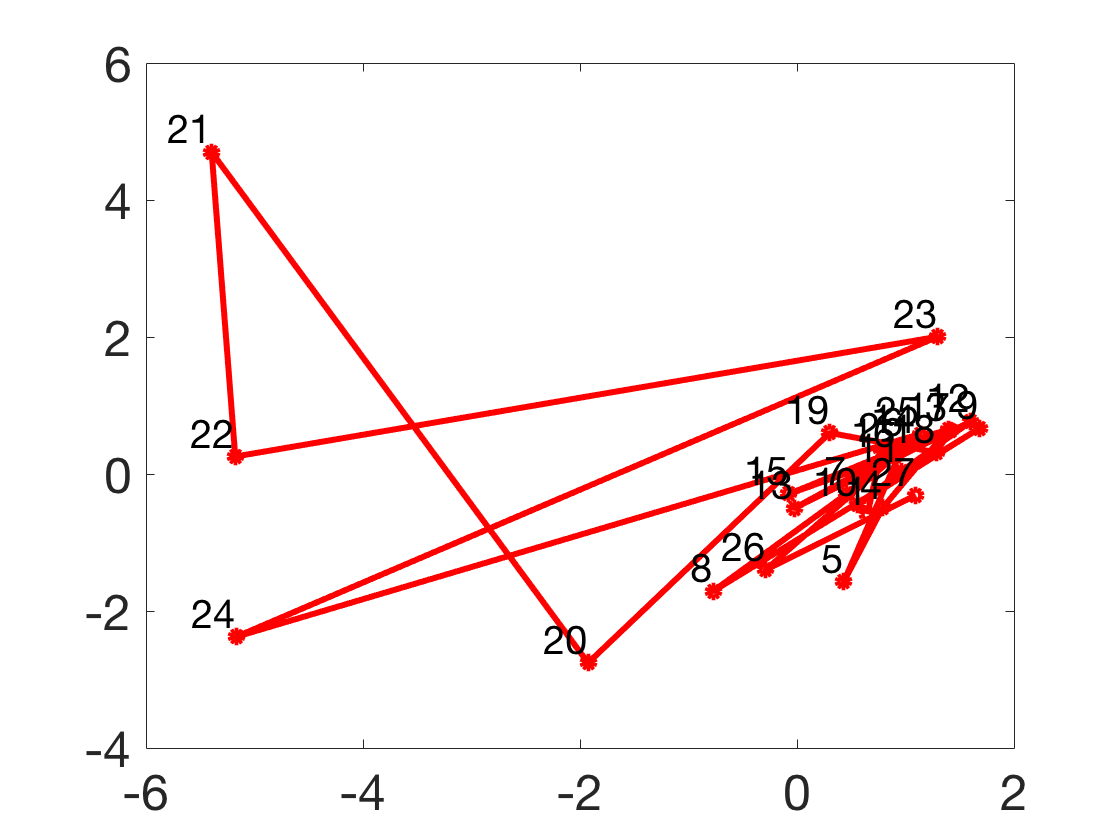}
	\end{tabular}
	\vspace{-0.3cm}
	\caption{Two-dimensional visualization of the path in shape space using isomap \cite{tenenbaum2000global} dimensionality reduction to project the path of shapes after shape filter on the Riemannian manifold. From left to right columns: the values for $\rho$ are 0.01, 0.2, 0.4, 0.6, 0.8 ,1. Top row: isomaps for using Bi3 local; bottom row: isomaps for using sGaussian. The markers on the spline shows the indices of time. Note that the scales for of each diagram from left to right is getting larger.}
	\label{fig:iso}
	\vspace{-0.4cm}
\end{figure*}

We evaluate the performance of these three weights using the \textit{Dice coefficient} and \textit{normalized mean squared error}. Table. \ref{tbl:results} shows the mean scores of 20 experiments. The Dice coefficient (Dice $\in[0,1]$) compares the similarity between two sets: the ground truth segments $S_g$ and the filtered segments $S_f$. They are both 3D volumes with packed 2D time-indexed results. Dice is calculated by $\frac{2|\mathrm{S_g} \cap \mathrm{S_f}|}{|\mathrm{S_g}|+|\mathrm{S_f}|}$ , where $|\cdot|$ denotes the cardinality of the corresponding set. The mean squared error ($\text{MSE}=\frac{1}{Z}\|\mathrm{S_g}-\mathrm{S_f}\|_{2}^{2}$) measures the average squared errors between ground truth and test data. The MSE is normalized by the total number of pixels $(Z)$ of all the frames. By comparing Dice and MSE factors, we demonstrate the superiority of  Bi3 local and sGaussian over the unity weighting model. 


More intuitively, Fig. 3 visualizes the comparison of the paths before shape filter and after shape filters in Cartesian coordinates. These results show that the application of shape filter is twofold: 1) smoothing the non-smooth boundaries of preliminary results; 2) filtering out the extra neuron from the single neuron segmentation path by substituting a new "reasonable" shape. This figure also compares the filtered shapes with different weights graphically. Fig. 3 shows the comparison between data and smoothness oriented results based on the choice of $\rho$ in equation (\ref{eq:main}). To precisely verify the influence of $\rho$ on filtered path, we utilized the isomap dimensionality reduction method \cite{tenenbaum2000global} to project the high dimensional path on the Riemannian manifold to trajectories on a 2D Cartesian map as shown in Fig. 4. When $\rho$ is near 0, the projected trajectories become smoother and denser. In this case, using shape filter can result into a smooth shape regression on the manifold.

Overall, the proposed algorithm is computationally efficient. The major time consumption is the local weighting calculation with roughly $\mathcal{O}(2nt)$ computation complexity ($n$: number of sampling points; $t$: number of time-indexed data). The performance of shape filter will degrade if the quantity of input shapes is not sufficient to generate the path on the manifold that predicts the evolution of activated neurons.

\vspace{-0.5cm}
\section{Conclusion}
\vspace{-0.3cm}
The proposed SRV-based manifold shape filter provides a powerful theoretical basis for repairing time sequences of shapes obtained in calcium imaging of neurons. The method is agnostic to the initial segmentation even for the images depicting overlapping neurons. The experimental results demonstrate that the proposed weighting methods outperform the method with equal weights. Our shape filter is effective in detecting improper segmentation and addressing the splitting problem.




\bibliographystyle{IEEEbib}
\small
\bibliography{ShapeFilter}

\end{document}